\documentclass[aps,prb,reprint,amssymb,amsmath,superscriptaddress]{revtex4-1}
\usepackage{graphicx}
\usepackage{epstopdf}
\input{ulem.sty}

\usepackage[english]{babel}
\bibpunct{[}{]}{,}{n}{}{}
\begin{document}

\title{Magneto-intersubband oscillations in two dimensional systems  with energy spectrum split due to spin-orbit interaction}

\author{G. M.~Minkov}

\affiliation{School of Natural Sciences and Mathematics, Ural Federal University,
620002 Ekaterinburg, Russia}

\author{O.\,E.~Rut}
\affiliation{School of Natural Sciences and Mathematics, Ural Federal University,
620002 Ekaterinburg, Russia}

\author{A.\,A.~Sherstobitov}

\affiliation{School of Natural Sciences and Mathematics, Ural Federal University,
620002 Ekaterinburg, Russia}

\affiliation{M.~N.~Miheev Institute of Metal Physics of Ural Branch of
Russian Academy of Sciences, 620137 Ekaterinburg, Russia}

\author{A.\,V.~Germanenko}
\affiliation{School of Natural Sciences and Mathematics, Ural Federal University,
620002 Ekaterinburg, Russia}

\author{S.\,A.~Dvoretski}

\affiliation{Institute of Semiconductor Physics RAS, 630090
Novosibirsk, Russia}

\author{N.\,N.~Mikhailov}

\affiliation{Institute of Semiconductor Physics RAS, 630090
Novosibirsk, Russia}
\affiliation{Department of Physics, Novosibirsk State University, Novosibirsk 630090, Russia}

\author{S.\,V.~Ivanov}

\affiliation{Ioffe Physical Technical Institute,  St. Petersburg 194021, Russia}

\author{V.\,A.~Soloviev}

\affiliation{Ioffe Physical Technical Institute,  St. Petersburg 194021, Russia}

\author{M.\,U.~Chernov}

\affiliation{Ioffe Physical Technical Institute,  St. Petersburg 194021, Russia}
\date{\today}

\begin{abstract}
In the present paper we study magneto-intersubband oscillations (MISO) in HgTe/Hg$_{1-x}$Cd$_x$Te single quantum well  with ``inverted''  and ``normal'' spectra and in conventional In$_{1-x}$Ga$_x$As/In$_{1-y}$Al$_y$As quantum wells with normal band ordering. For all the cases when two branches of the spectrum arise due to spin-orbit splitting, the mutual arrangement of the antinodes of the Shubnikov-de Haas oscillations and the maxima of  MISO occurs opposite to that observed in  double quantum wells and in  wide quantum wells with two subbands occupied and does not agree with the  theoretical predictions. A ``toy'' model is supposed that  explain qualitatively this unusual result.
\end{abstract}


\maketitle

\section{Introduction}
\label{sec:intr}

The magnetic field ($B$) normal to the plane of a two-dimensional
gas is responsible for the Landau quantization of
the spectrum and, as a result, for oscillations of the transverse resistivity ($\rho_{xx}$) at low temperatures, known as the Shubnikov-de Haas (SdH) oscillations.  The oscillating part of $\rho_{xx}$ is given by the well known expression \cite{LifKos55}
\begin{equation}
\label{eq10}
\frac{\Delta\rho^{SdH}}{\rho_D}=2 \frac{\Delta\nu}{\nu_0}\mathcal{F}\left(\frac{2\pi^2k_BT}{\hbar\omega_c}\right),
\end{equation}
where $\rho_D$ and $\nu_0$  stand for  resistivity and   density of states in zero magnetic field, respectively,   $\mathcal{F}(x)=x/\sinh{x}$, $\omega_c=eB/m$, $m$ is the effective mass, and
\begin{equation}
\label{eq15}
 \frac{\Delta\nu}{\nu_0}=-2\delta\, \cos{\left(\frac{2\pi f}{B}\right)}
\end{equation}
with  $\delta=\exp{\left(-2\pi\gamma/\hbar\omega_c\right)}$, where $\gamma$ is the broadening of the Landau levels. As seen the SdH oscillations are periodical in the reciprocal magnetic field and their frequency is determined by the electron density ($n$) and the degree of degeneracy of the Landau levels ($s$); $f=n/s \times 2\pi\hbar /e$ (see, e.g., Ref.~\cite{Coleridge89}).

In systems in which two branches of the energy spectrum are occupied, in
addition to oscillations with the frequencies $f_1$ and $f_2$ determined by the electron densities in these branches $n_1$ and $n_2$, the oscillations at the difference frequency ($f_1-f_2$) appear due to transitions between these branches (see \cite{Dmitriev12,Polyanovs88,Leadly92} and references therein). They are named magneto-intersubband oscillations (MISO). These two  branches can be, e.g., the two  subbands of spatial quantization in a relatively wide quantum well (QW) or two subbands belonging to different quantum wells in a double quantum well heterostructures. Such
oscillations were widely studied both theoretically \cite{Mamani08,Averkiev01-1,Raikh94,Raichev10} and experimentally
in various semiconductor structures \cite{Sander98,Mamani09,Mamani09-1,Wiedmann10,Col90,Bykov19}.
Theoretically they can be described by the following expression \cite{Mamani09-1}
\begin{equation}
\label{eq80}
\frac{\Delta\rho^\text{MISO}_{xx}}{\rho_D}=\delta_1\delta_2\frac{1}{\tau_{12}}
2\frac{n_1\tau_1+n_2\tau_2}{n_1+n_2}\cos{\left[\frac{2\pi(f_1-f_2)}{B}\right]},
\end{equation}
where $1/\tau_{12}=W_{12}$ is the probability of transitions between states of different branches averaged over scattering angles, and $1/\tau_{1}$ and $1/\tau_{2}$ are the scattering rates wich includes both intrasubband and intersubband scattering for the branches $1$ and $2$, respectively, $\rho_D=(\rho_1^{-1}+\rho_2^{-1})^{-1}$, and $\rho_1=m_1/e^2n_1\tau_1$, $\rho_2=m_2/e^2n_2\tau_2$.

Thus, the total resistivity oscillations are as follows
\begin{equation}
\label{eq90}
\frac{\Delta\rho}{\rho_D}=\frac{\Delta\rho^{SdH}_1}{\rho_1}\frac{\rho_D}{\rho_1}+
\frac{\Delta\rho^{SdH}_2}{\rho_2}\frac{\rho_D}{\rho_2} + \frac{\Delta\rho^\text{MISO}_{xx}}{\rho_D}.
\end{equation}
In the case of $\gamma_1\simeq \gamma_2=\gamma$,  $m_1\simeq m_2=m$, and $\rho_1\simeq\rho_2=\rho$ the sum of the first two terms can be represented in the form
\begin{eqnarray}
\label{eq95}
\frac{\Delta\rho^{SdH}}{\rho_D}&=&\frac{\Delta\rho^{SdH}_1}{2\rho}+
\frac{\Delta\rho^{SdH}_2}{2\rho} \nonumber \\
&\simeq& -\mathcal{F}\,\delta\left[
\cos{\left(\frac{2\pi f_1}{B}\right)}+\cos{\left(\frac{2\pi f_2 }{B}\right)}\right]
\nonumber \\
&=& -2\mathcal{F}\,\delta \cos{\left[\frac{\pi(f_1-f_2)}{B}\right]} \cos{\left[\frac{\pi(f_1+f_2)}{B}\right]},
\end{eqnarray}
which  at $f_1\neq f_2$ describes the beatings of the SdH oscillations.

Expressions (\ref{eq80}) and (\ref{eq95})  show that the magnetic fields corresponding to antinodes of SdH oscillations should coincide with the magnetic fields of the maxima of MISO.
\begin{figure}
\includegraphics[width=\linewidth,clip=true]{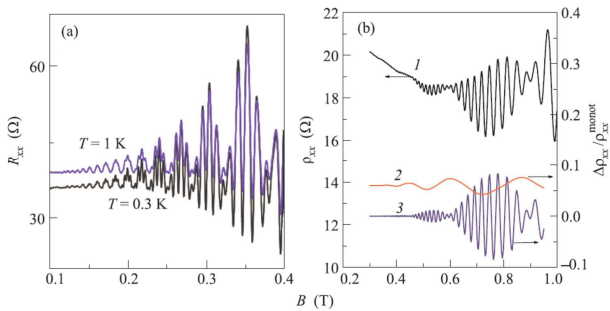}
\caption{(Color online) (a) The magnetoresistance of the GaAs double quantum well with $d_{QW}=14$~nm and $d_{barrier}=1.4$~nm from
Ref.~\cite{Mamani08}. (b) The magnetoresistance of the HgTe single quantum well of  $10$~nm width grown on (013) GaAs substrate  for $n=8.7\times 10^{11}$~cm$^{-2}$, $T= 4$~K from
Ref.~\cite{Minkov19}. The curve 1 is experimental, the curves 2 and 3 are the low- and  high-frequency components, respectively, obtained by the decomposition of the experimental dependence 1 as described in Section~\ref{sec:exp0}.}
\label{F10}
\end{figure}

Namely  such a mutual position of MISO and antinodes  is observed in all the cases mentioned above. As an example  we show the magnetoresistance oscillations  in double quantum well heterostructure from Ref.~\cite{Mamani08} in  Fig.~\ref{F10}(a). It is clearly seen that the low-frequency oscillations (these are MISO) are observed in low fields. With an increase in the magnetic field, high-frequency SdH oscillations with amplitude beats are superimposed on these oscillations, and the antinodes of the  high-frequency oscillations correspond to the  MISO maxima.

However, recently \cite{Minkov19} an opposite mutual position of the antinodes of the SdH and MISO oscillations was observed in HgTe single QW with a width of $10$~nm grown on substrate of (013) orientation [see Fig.~\ref{F10}(b)]. In general, they have only one important difference from those shown in Fig.~\ref{F10}(a); the magnetic fields corresponding to the antinodes of the high-frequency oscillations correspond to the minima of MISO. For clarity, in Fig.~\ref{F10}(b) we show separately the low- and high-frequency oscillations (lower curves) obtained by the decomposition of the experimental curve (the procedure of decomposition and analysis are described below).

The quantum wells in heterostructures HgTe/Cd$_x$Hg$_{1-x}$Te have a number of unusual properties compared to quantum wells based on semiconductors with non-zero band gap. The reason for this is inverted order of $\Gamma_6$ and $\Gamma_8$ band in bulk HgTe in which $\Gamma_6$ band, forming a conduction band in conventional semiconductors, is located below the $\Gamma_8$ band, which forms the valence band. Such arrangement  leads to features of the energy spectrum of two-dimensional carriers, knowledge of which is required for reliable interpretation of all phenomena in structures with HgTe quantum wells.

The energy spectrum of HgTe quantum wells is complicated and depends strongly on the quantum well width ($d$). For $d<d_c\simeq6.3$~nm, the conduction band is formed from electron states and the states of the light hole \cite{Gerchikov90,Zhang01,Novik05,Bernevig06,ZholudevPhD}. This type of the spectrum is named ``normal''. At $d>d_c$, the conduction band is formed from the heavy-hole states and such type of the spectrum is named ``inverted'' \footnote{It should be clarified here. The terms \textit{inverted} and \textit{normal} with respect to the spectrum of quantum wells of gapless semiconductors differ from the similar terms \textit{inverted} and \textit{normal order} of the bands of bulk semiconductors. In the bulk semiconductors, the term \textit{normal} means that the upper in energy is the doubly degenerate $\Gamma_6$ band (conduction band), while the valence band is formed by the fourfold degenerate $\Gamma_8$ states. The term \textit{inverted} means that the $\Gamma_8$ band is located higher in energy than the $\Gamma_6$ band and one of its branches forms the conduction band. Therefore, regarding the spectrum of quantum wells, we will use quotation marks: ``normal'' and ``inverted''.}.

Let us return to the features of MISO in HgTe QW, shown in Fig.~\ref{F10}(b). Besides the fact that the MISO which is shown in this figure was observed in
a quantum well with an ``inverted'' spectrum, two branches of the spectrum in
it appeared due to strong spin-orbit (SO) splitting \cite{Minkov19}. Magneto-intersubband oscillations for the case of strong SO coupling were theoretically considered  in Refs.~\cite{Langenbuch04,Novoksh}. There was obtained that the magnetic fields corresponding to antinodes of SdH oscillations should coincide with the  maxima position of MISO also.

Thus, the reason that leads to the unusual mutual position of the antinodes of the SdH oscillations and the MISO remains unclear. This may be a low symmetry of the substrate, the ``inverted'' spectrum, a gapless spectrum of the parent material of the quantum well.

In this paper we present the results of the experimental investigations of the  magneto-intersubband oscillations in HgTe/Hg$_{1-x}$Cd$_x$Te single quantum well  with the ``inverted''  and ``normal'' spectra and in conventional In$_{1-x}$Ga$_x$As/In$_{1-y}$Al$_y$As quantum wells with normal band ordering.

The paper is organized as follows. The samples and experimental conditions are described in the next section. The experimental results and their analysis for the HgTe and In$_{1-x}$Ga${_x}$As based QWs are presented in Sections~\ref{sec:exp0} and \ref{nAlAs/InGaAS}, respectively. Section~\ref{discus} is devoted to the discussion of possible reasons for unusual mutual position of antinodes and maxima of MISO and interpretation of the data obtained within the framework of the simple phenomenological  model.  The conclusions are given in Section~\ref{conlc}.

\section{Experimental}
\label{sec:expdet}

Our samples with the HgTe quantum wells  were realized on the basis of
HgTe/Hg$_{1-x}$Cd$_{x}$Te ($x=0.39-0.6$) heterostructures grown by the
molecular beam epitaxy on a GaAs substrate with the different surface
orientations \cite{Mikhailov06}. The sample with  In$_{0.75}$Ga$_{0.25}$As/In$_{0.75}$Al$_{0.25}$As quantum well was grown on the (001) GaAs semi-insulating substrate. The parameters of the structures under study are presented in the Table~\ref{tab1}.

\begin{table}
\caption{The parameters of  heterostructures under study }
\label{tab1}
\begin{ruledtabular}
\begin{tabular}{ccccccc}
 \#& QW material & structure & $d$ (nm)& $n\footnotemark[1]$
(cm$^{-2}$) &   orientation \\
\colrule
  1& HgTe & 100623 & $18$  & $1.15\times 10^{11}$   &
$(100)$\\
  2& HgTe & 091228 & $14$   & $1.6\times 10^{11}$ &    $(211)$\\
  3& HgTe & 150220 & $4.6$    & $1.6\times 10^{11}$ &
$(013)$\\
  4& InGaAs & A1-849 & 30  &  $3.5\times 10^{11}$ &
$(100)$\\
\end{tabular}
\end{ruledtabular}
\footnotetext[1]{At $V_g=0.$}
\end{table}

The samples were mesa etched into standard Hall bars of $0.5$~mm  width and the distance between the potential probes was $0.5$~mm. To change and control the carrier density in the quantum well, the field-effect transistors were fabricated with parylene as an insulator and aluminium as a gate electrode. For each heterostructure, several samples were fabricated and studied. All measurements were carried out using the  DC technique in the linear response regime at $T=(1.3\ldots 10.0)$~K within the magnetic field range $(-2.0\ldots 2.0)$~T.

\begin{figure}
\includegraphics[width=\linewidth,clip=true]{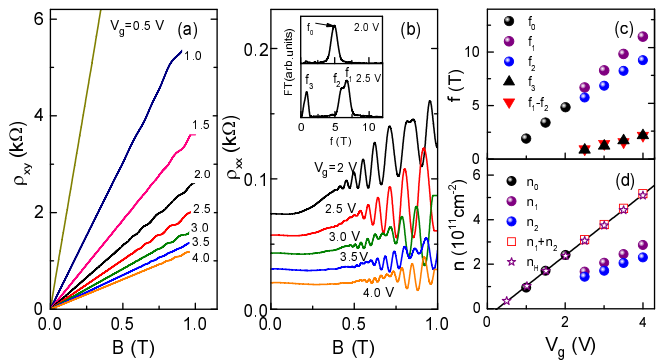}
\caption{(Color online) (a) and (b) The magnetic field dependences of $\rho_{xy}$ and $\rho_{xx}$, respectively, for the structure 1 with substrate orientation (100) at some gate voltages $V_g$, $T=4$~K. Each curve for $V_g< 4.0$~V in panel (b) is shifted up by the value of $0.01$~k$\Omega$ relative to the previous curve. The inset in (b) shows the Fourier spectrum of $\rho_{xx}(B^{-1})$ for the two gate voltages. (c) The gate voltage dependence of the Fourier maxima positions. (d) The gate voltage dependences of the electron densities in split subbands and the total electron density obtained as described in the text.}
\label{F20}
\end{figure}

\section{MISO in the mercury telluride quantum wells}
\label{sec:exp0}
We begin our analysis with the results obtained for the structure 1. This
structure has an inverted spectrum, but unlike the structure, the results for which are shown in Fig.~\ref{F10}(b), it was grown on a substrate with orientation (100). To characterize the structure, the magnetic field dependences of $\rho_{xy}$ and $\rho_{xx}$ for some gate voltages are presented in Figs.~\ref{F20}(a) and \ref{F20}(b), respectively.  As seen the dependences $\rho_{xy}(B)$  are linear   for $B<0.5$~T, and  the Hall density $n_H=-1/eR_H(0.2~\text{T})$ increases linearly with increasing $V_g$ as shown in Fig.~\ref{F20}(d).  The oscillations of $\rho_{xx}(B)$ are visible starting from $B\simeq (0.3-0.4)$~T. The Fourier analysis of the oscillations shows that the oscillations of only one frequency $f_0$ are observed  at $V_g<2$~V [see the inset in Fig.~\ref{F20}(b) and Fig.~\ref{F20}(c)]. The electron density found from this frequency under assumption of twofold degeneracy of the Landau levels  $n_0=f_0\times e/\pi\hbar$  coincides with the Hall density [see  Fig.~\ref{F20}(d)].

\begin{figure}
\includegraphics[width=0.9\linewidth,clip=true]{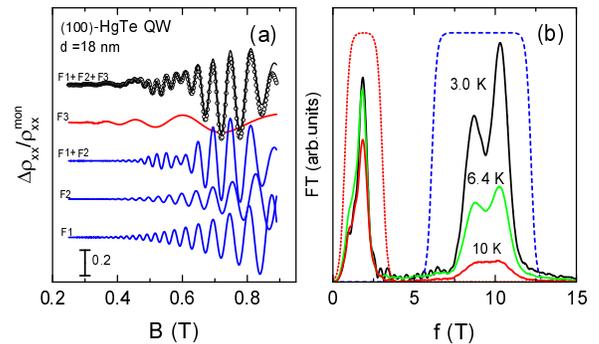}
\caption{(Color online) (a) The oscillating part of the $\rho_{xx}(B)$ for the structure 1 for $V_g=3.5$~V, $T=3.0$~K (circles). The electron density and mobility are equal to $n_H=4.5\times 10^{11}$~cm$^{-2}$ and $\mu=4.6\times 10^{5}$~cm$^2$/V$\cdot$s, respectively. The curves are the  results of data analysis (see the text). (b) The Fourier spectra of the oscillations for the different temperatures (the solid lines) and the bandpass filters (the dotted and dashed lines) used to separate the low- and high-frequency components.}
\label{F30}
\end{figure}

With the growing  gate voltage, at $V_g>2$~V, the  beating of the high-frequency oscillations and the appearance of the low-frequency oscillations are observed.  Therewith the Fourier spectra exhibit three components with the frequencies $f_1$, $f_2$, and $f_3$ [see the inset in Fig.~\ref{F20}(b) and Fig.~\ref{F20}(c)]. They can be much better resolved if one treats the oscillating part of resistivity $\Delta\rho_{xx}(B)=\rho_{xx}(B)-\rho^{mon}_{xx}(B)$, where $\rho^{mon}_{xx}(B)$ is the monotonic part of $\rho_{xx}(B)$. The written is illustrated by Fig.~\ref{F30}(a) in which we plot $\Delta\rho_{xx}(B)/\rho^{mon}_{xx}(B)$ for $V_g=3.5$~V ($n_H=4.5\times 10^{11}$~cm$^{-2}$) and by Fig.~\ref{F30}(b) which shows the Fourier spectra of these oscillations obtained in the low magnetic field, $B<0.85$~T, before onset of the quantum Hall effect regime.

The sum of densities $n_1$ and $n_2$ found  from  $f_1$ and $f_2$  under assumption of nondegeneracy of Landau levels ($n_{1, 2}=f_{1, 2}\times e/2\pi\hbar$), shown by squares in Fig.~\ref{F20}(d)  coincides with the Hall density. This is clear indication of the fact that the splitting of the Fourier spectra  observed at $n>2.5\times 10^{11}$ cm$^{-2}$ is a consequence of the SO splitting of the spectrum, and $n_1$ and $n_2$ are the electron densities in the  split subbands.

The origin of low-frequency oscillations with the frequency $f_3$ becomes clear from the  Fig.~\ref{F20}(c), which demonstrates that $f_3$ is equal to $f_1-f_2$. This shows that the low-frequency oscillations are a consequence of intersubband transitions. This conclusion is also supported by the temperature dependence of the amplitude of these oscillations, which, as the theory predicts, decreases significantly slower with the increasing temperature than the amplitudes of the oscillations with the frequencies $f_1$ and $f_2$ [see Fig.~\ref{F30}(b)].

Thus, MISO in our structure arise due to transition between two single-``spin'' branches of the energy spectrum split due to the  SO interaction.

To determine the frequencies and amplitudes of the different oscillation components more accurately, we used   the bandpass filtering as shown in Fig.~\ref{F30}(b). Then,  applying the inverse Fourier transformation we obtained  the oscillations corresponding to $f_3$  frequency [the curve F3 in Fig.~\ref{F30}(a)] and superposition of oscillations with two higher frequencies $f_1$ and $f_2$  [the curve labeled as F1+F2 in Fig.~\ref{F30}(a)]. Finally, the oscillation curve  F1+F2  was fitted to the sum of two Lifshits-Kosevich formulas \cite{LifKos55};
\begin{eqnarray}
\label{eq96}
\frac{\Delta\rho}{\rho}&=&\sum_{i=1}^2\beta_i \exp{\left(-\frac{2\pi\gamma_i}{\hbar\omega_c^i}\right)} \nonumber \\ &\times&\mathcal{F}\left(\frac{2\pi^2k_BT}{\hbar\omega_c^i}\right)\cos{\left(\frac{2\pi f_i}{B}+\phi_i\right)}.
\end{eqnarray}
The fitting parameters were two frequencies  determined by the electron densities $f_i=n_i \times 2\pi\hbar /e$, two damping parameters $\gamma_i$, two prefactors $\beta_i$, and two phases of oscillations $\phi_i$. The quality of the fitting procedure is illustrated by Fig.~\ref{F30}(a). Although this fitting procedure involves such a large number of the fitting parameters, it gives an unambiguous result on the electron densities $n_1$ and $n_2$. The $n_1$ and $n_2$ values found with this data treatment are shown  in Fig.~\ref{F20}(d) by the balls.

\begin{figure}
\includegraphics[width=0.9\linewidth,clip=true]{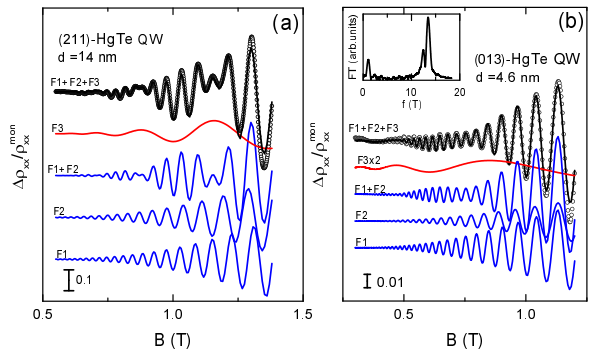}
\caption{(Color online) (a) Oscillation part of $\rho_{xx}$ (circles) for the structure 2 with substrate orientation (211) and the results of the  data analysis, $n_H=8.3\times 10^{11}$~cm$^{-2}$, $\mu=3.6\times 10^{5}$~cm$^2$/V$\cdot$s. (b) The analogous data as in the panel (a) for the structure 3 with $d=4.6$~nm$<d_c$, $n_H=6.4\times 10^{11}$~cm$^{-2}$,  $\mu=0.48\times 10^{5}$~cm$^2$/V$\cdot$s.  The inset shows the Fourier spectrum of the oscillations. }
\label{F40}
\end{figure}

As seen in Fig.~\ref{F30}(a), in the structure 1 with the orientation $(100)$,  the magnetic fields of antinodes  of the high-frequency oscillations coincide with the MISO minima, just as in the structure with the orientation (013) which oscillations are shown in Fig.~\ref{F10}(b) \cite{Minkov19}. Such a mutual position of antinodes and MISO minima is observed for all the electron densities, where SO splitting manifests itself well. Analogous results were obtained for the  structure 2 with orientation (211) [see Fig.~\ref{F40}(a)].

Thus, in all the structures with ``inverted'' spectrum ($d>d_c$) with different orientations, (100), (211), and  (013), the antinode positions of SdH oscillations coincide with that of the MISO minima. The question arises: is this not due to the fact that the spectrum is inverted in these structures?

To find out this, a structure with the ``normal'' spectrum was studied also. The results obtained  by the same data processing in structure 3 with $d=4.6~\text{nm}<d_c$ are presented in Fig.~\ref{F40}(b). The beating of oscillations is clearly evident in this case also, however, the amplitude in the nodes remains quite large. This is due to the fact that the amplitudes of the two high-frequency oscillations  noticeably differ [see the inset and compare the curves F1 and F2 in Fig.~\ref{F40}(b)]. Nevertheless, it can be argued that in this structure the positions of antinodes in magnetic field are close to minima of MISO again.

Therefore, to understand whether the MISO feature in the HgTe quantum wells is related to the feature of the spectra HgTe wells or to the fact that two
branches of the spectrum arise due to spin-orbit  splitting, it is useful to
study MISO in a  structure based on the semiconductor with a normal band ordering.

\section{ MISO in   indium-gallium arsenide quantum well}
\label{nAlAs/InGaAS}

\begin{figure}
\includegraphics[width=0.9\linewidth,clip=true]{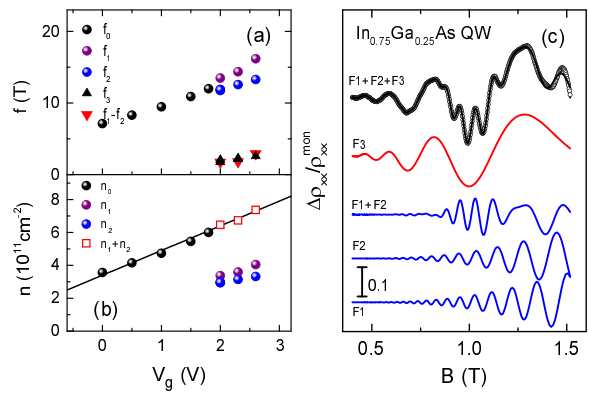}
\caption{(Color online) (a) and (b) The gate voltage dependences of the Fourier maxima positions and oscillation frequencies and electron densities, respectively, found from SdH oscillations as described in Section~\ref{sec:exp0} for  In$_{0.75}$Ga$_{0.25}$As QW (the sample 4), $T=1.3$~K. (c) The oscillation part of the magnetoresistance (the circles) and its decomposition. $n_H=6.2\times 10^{11}$~cm$^{-2}$,  $\mu=1.67\times 10^{5}$~cm$^2$/V$\cdot$s.}
\label{F50}
\end{figure}

In order to verify that the discussed mutual positions of the nodes in the SdH oscillations and maxima of  MISO does not relate  to the features of the energy spectrum of the quantum wells of the gapless HgTe semiconductor, we studied the QW of the narrow-gap In$_{0.75}$Ga$_{0.25}$As semiconductor.  The bulk In$_{0.75}$Ga$_{0.25}$As has normal order of the bands with $E_g= E_{\Gamma_6}(k=0)-E_{\Gamma_8}(k=0)\simeq  0.7$~eV. The gate voltage dependences of the oscillation frequencies and electron densities are shown in Figs.~\ref{F50}(a) and \ref{F50}(b), respectively.  As evident  the splitting appears at $V_g\simeq 2.0$~V, when the electron density reaches the value  $\simeq6\times 10^{11}$~cm$^{-2}$. The resistivity oscillations for  $V_g=2.25$~V   are shown in Fig.~\ref{F50}(c) by circles. It is clearly seen that the $\rho_{xx}$ oscillations are superposition of the oscillations with low and high frequencies. Further data processing and analysis were the same as described in Section~\ref{sec:exp0}. The results are shown in Fig.~\ref{F50}(c). It is seen that there are one low-frequency (the curve F3) and two high-frequency components (the curves F1 and F2). The sum of the last two leads to the beating  (the curve labeled as F1+F2). Again,   the positions of the antinodes of high-frequency oscillations coincide with the minima of the $\rho_{xx}$ MISO.

Thus, in the structure based on the conventional  semiconductors with normal band ordering, in which two branches of the spectrum are formed due to SO interaction, the positions of the antinodes of high-frequency oscillations also coincide with the minima of the magneto-intersubband oscillations.

\section{ Discussion}
\label{discus}

The above results show that the relative positions of antinodes and of the MISO maxima in the case when the two energy spectrum branches  arise as a result of SO splitting are opposite to that observed in double quantum well structures and in  wide quantum wells regardless of the orientation HgTe QW, the type of spectrum (``inverted'' or ``normal'', i.e., $d>d_c$ or $d<d_c$), the parent materials (normal In$_{1-x}$Ga$_{x}$As or gapless HgTe), and do not agree with that predicted  theoretically, Eqs.~(\ref{eq10})--(\ref{eq90}).

Let us consider which approximations were used to obtain  Eqs.~(\ref{eq10})--(\ref{eq90}). It was assumed that the magnetic field dependence of the energy of the Landau levels is described by a simple semiclassical formula, i.e., the Berry phase is equal to zero. In the  structures with a complex energy spectrum, this may be not so. It would seem that the procedure used to separate the contributions of each branch of the spectrum to the oscillations gives their phase values  also. However, the  cosine arguments in Eq.~(\ref{eq96}) are determined  accurately, but for a large number of Landau levels (i.e., before onset of the quantum Hall effect), the accuracy in the phases $\phi_1$ and $\phi_2$  is quit low.

\begin{figure}
\includegraphics[width=0.7\linewidth,clip=true]{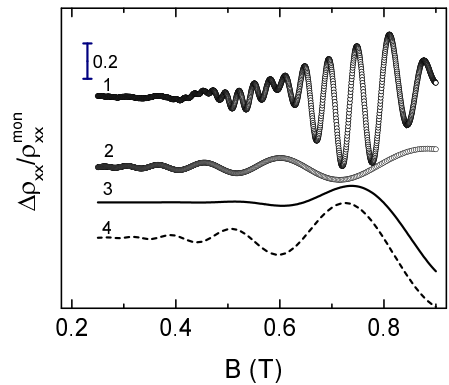}
\caption{ (Color online) Oscillating part of $\rho_{xx}(B)$ (the curve 1) and MISO (the curve 2) in the structure  1 [they are  the same curves as in Fig.~\ref{F30}(a)]. The curve 3  is the low-frequency part of the product of the curves 1 and 2 from Fig.~\ref{F30}(a).  The curve 4 is the curve 2 multiplied by a factor  $0.04\exp(2.9/B)$ in order to make the low-field oscillations more visible.}
\label{F60}
\end{figure}

The most clear distinction between MISO in the structures studied and MISO  predicted by  Eqs.~(\ref{eq80}) and (\ref{eq95}) can be demonstrated as follows.  As shown in Refs.~\cite{Col90,Leadly92},  $\Delta\rho_{xx}^\text{MISO}(B)$ is proportional to the low-frequency part of the product of  the densities of states coming from the different branches and oscillating with the frequencies $f_1$ and $f_2$ by the probability
of transitions between these branches $W_{12}$.
We have shown above that the analysis of the experimental results allows us to obtain the component with frequencies $f_1$ and $f_2$ separately [see, e.g., curves F1 and F2 in Fig.~\ref{F50}(c)].  If we multiply these two components and apply the digital filtering we can obtain the low-frequency part of the product which can be compared with the experimental MISO. Here there are no any assumptions because we operate with the experimental data. An example of such a data processing is shown in  Fig.~\ref{F60}.  As clearly seen the low-frequency component of the product of $f_1$ and $f_2$ (the curves 3 and 4) is in antiphase with the MISO observed experimentally (the curve 2). This shows that antiphase results from a peculiarity of transition  rate but not from a peculiarity of the densities of states $\nu_1$ and $\nu_2$.

Naively, one could expect that the nontrivial values of the Berry phases could improve the situation by changing the mutual positions of the antinodes and MISO maxima. However, it is not the case. It easy to show that nonzero $\phi_1$ and $\phi_2$ values, resulting in a shift of the antinodes  on the value $(\phi_1-\phi_2)/2$ in a reciprocal magnetic field, simultaneously  result in the phase shift of  the MISO maxima on the value $(\phi_1-\phi_2)$. Thus, the Berry phases are not the cause of the unusual mutual positions of the nodes and maxima of $\rho_{xx}$ MISO.

Another approximation made is the neglect of the dependence of probability of the transitions between the $i$-th and $j$-th Landau levels of different branches ($W_{ij}$)  on the detuning from resonance, i.e., on the difference  between their energies $\Delta_{ij}=E_i^{(1)}-E_j^{(2)}$. For this reason the transition rate between Landau levels of the different branches equal to $W_{ij}\nu^{(1)}(E_i^{(1)},B)\nu^{(2)}(E_j^{(2)},B)$ is maximal in resonance, i.e., when $E_i^{(1)}=E_i^{(2)}=E_F$ [here, $\nu^{(1)}(E_i^{(1)},B)$ and $\nu^{(2)}(E_j^{(2)},B)$ are the densities of states  of Landau levels of different branches]. In our case, when the  two branches arise due to a strong SO coupling,  $W_{ij}$ can be sensitive to the detuning energy $\Delta_{ij}$. Indeed, the ``spins'' that are different in different branches are locked with the momentum of the corresponding branch, and the electron, which transit from the Landau level of one branch to a Landau level of the other one should change both ``spin'' and ``momentum''. The probability of such transitions in the absence of magnetic impurities can be markedly suppressed in resonance.

To assess the consequences of existence of the dependence of $W_{ij}$ on $\Delta_{ij}$ in the  positions of  MISO maxima we consider a purely phenomenological ``toy'' model. We assume that  $W_{ij}$ depends on the energy difference $\Delta_{ij}$ as
\begin{equation}\label{eq98}
  W_{ij}=\frac{1}{\tau_{12}}\left(1-h\frac{b^2}{\Delta_{ij}^2+b^2}\right).
\end{equation}
How, at least qualitatively, can the role of this effect be taken into account? It cannot be taken into account in the commonly used expression, Eq.~(\ref{eq80}), since it was obtained as a result of summing over the Landau levels under the assumption that $W_{ij}$ does not depend on the difference in their energies.

In order to take into account the dependence of $W_{ij}$ on $\Delta_{ij}$, one should return  to the original expression for the $\Delta\rho_{xx}^\text{MISO}$ namely to summing  over the Landau levels. As already mentioned above the oscillation part of $\rho_{xx}^\text{MISO}(B)$ is proportional to the low-frequency component (LFC) of  the product of the probability
$W_{ij}$ by the oscillating parts of the densities of the initial and final states, i.e., it can be written in the following form:
\begin{eqnarray}\label{eq100}
  \frac{\Delta\rho^\text{MISO}_{xx}}{\rho_D}&=&\frac{\mathcal{K}}{\nu_0^2} \sum_{i} \left\{\nu^{(1)}(E_{i}^{(1)},B)\right. \nonumber \\
  &\times & \left.  \sum_{j} \left.\nu^{(2)}(E_{j}^{(2)},B) W_{ij}(E_{i}^{(1)}-E_{j}^{(2)}) \right\}\right|_{\text{LFC}},
\end{eqnarray}
where $\nu(E,B)=eB/2\pi\hbar\times\gamma/\pi[(E-E_F)^2+\gamma^2]$ is the density of states of a  broadened  Landau level, $E_N^{(1),(2)}=\hbar\omega_c^{(1),(2)}\left(N+1/2\right)\pm\alpha
\sqrt{eB\left(N+1/2\right)/2\hbar}$ with $\alpha$ as the Rashba constant are the energies of the Landau levels, and $\mathcal{K}$ is some coefficient which has dimensions of time. In the specific case, when $W_{ij}$ is independent of energy ($W_{ij}=W_{12}=1/\tau_{12}$) $\mathcal{K}$ is
\begin{equation}\label{eq110}
\mathcal{K}=\frac{n_1\tau_1+n_2\tau_2}{n_1+n_2}
\end{equation}
and Eq.~(\ref{eq100}) coincides with Eq.~(\ref{eq80}) for $T=0$.

\begin{figure}
\includegraphics[width=0.9\linewidth,clip=true]{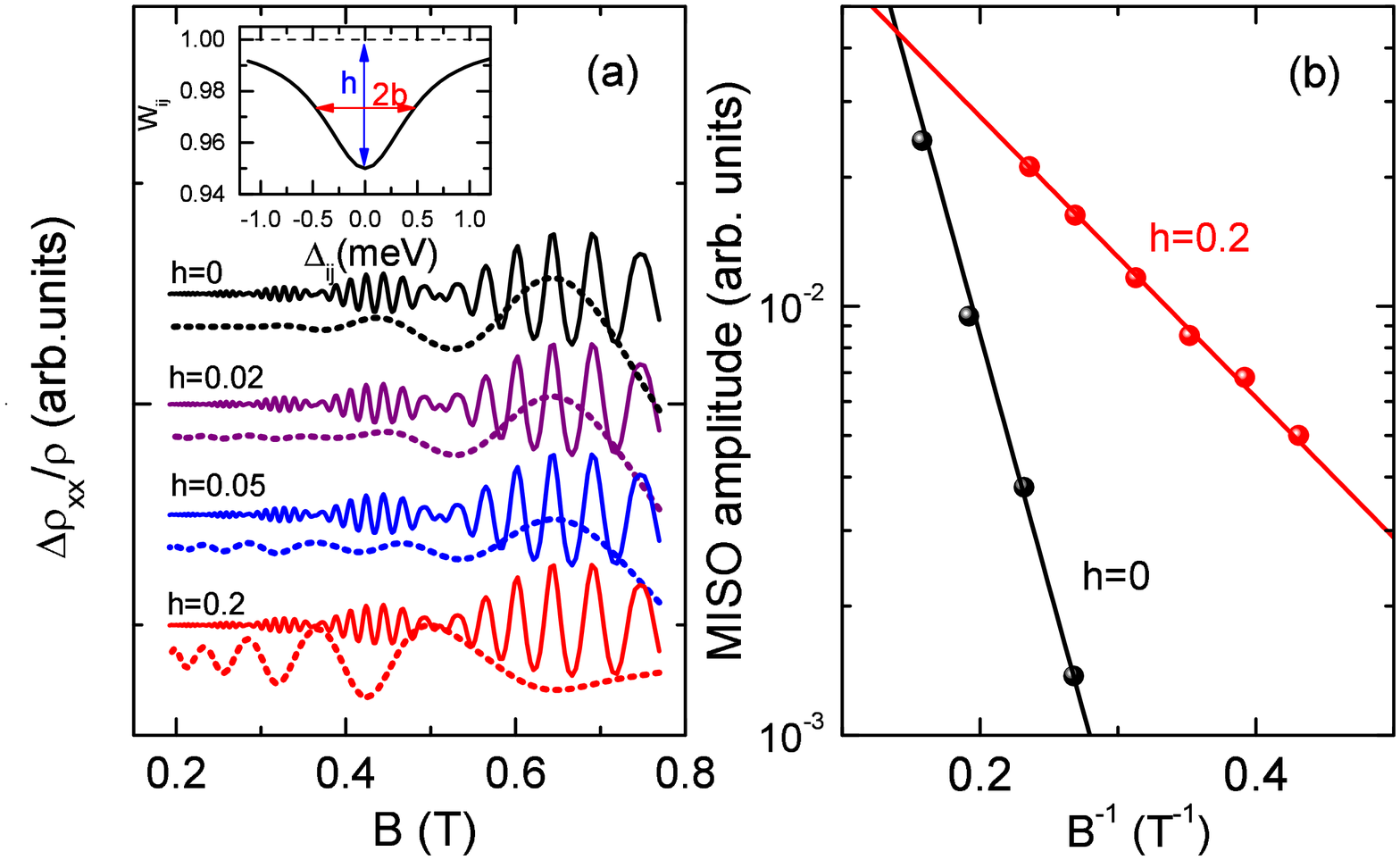}
\caption{(Color online)  (a) The results calculated within the framework of the ``toy'' model with the following set of parameters; $m=0.025 m_0$, $\alpha=3.5\times 10^{-6}$~meV$\cdot$cm, $b=0.5$~meV, and $E_F=42$~meV. The dashed lines are MISO calculated from Eq.~(\ref{eq100}) for the different $h$ values, the solid curves are the sum of the two components with the frequencies, corresponding to electron densities in spilt subbands  $n_1=2.5\times 10^{11}$~cm$^{-2}$ and $n_2=2.0\times 10^{11}$~cm$^{-2}$. (b) The $B^{-1}$ dependence of the MISO amplitudes for the two values of $h$ parameter.  }
\label{F70}
\end{figure}

In order to demonstrate how the dependence $W_{ij}(\Delta_{ij})$ changes the dependence $\Delta\rho^\text{MISO}_{xx}(B)$,  we present the $\Delta\rho^\text{MISO}_{xx}$ vs $B$ curves  calculated for several  $h$ values in Fig.~\ref{F70}(a). In the same figure the sum of the two high-frequency components  corresponding to the electron densities in split subbands are depicted also. When $h=0$, i.e., the expression Eq.~(\ref{eq100}) coincides with Eq.~(\ref{eq80}) for $T=0$, the positions of the antinodes correspond to the MISO maxima. At $h=0.02$,  the positions of the antinodes correspond to minima of $\Delta\rho_{xx}^\text{MISO}$ in low magnetic fields, $B<0.5$~T, but in the higher magnetic fields the maximum of $\Delta\rho_{xx}^\text{MISO}$ shifts and corresponds to the antinods again. At $h=0.2$, the positions of the antinodes correspond to minima of $\Delta\rho_{xx}^\text{MISO}$ over the whole $B$ range. Note the $B^{-1}$ dependence of the $\Delta\rho_{xx}^\text{MISO}$ amplitudes being exponential for any $h$ values weakens strongly with growing $h$ [see Fig.~\ref{F70}(b)]. The  significant weakening of the magnetic field dependence of the amplitude takes place at the presence of even a weak dependence of $W_{ij}$ on $\Delta_{ij}$   both for positive and for negative $h$ values.

\begin{figure}
\includegraphics[width=0.55\linewidth,clip=true]{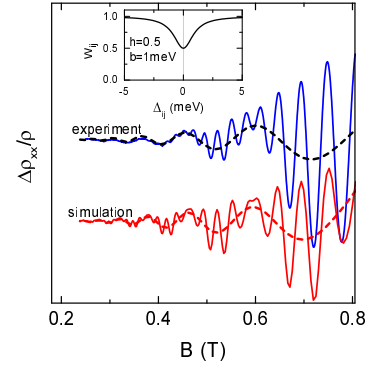}
\caption{The experimental (for the structure 1) and simulated dependences $\Delta\rho_{xx}(B)$ and $\Delta\rho_{xx}^\text{MISO}(B)$. The inset shows the   dependence $W_{ij}(\Delta_{ij})$ used in the simulating procedure. The experimental curve is divided by the temperature damping factor $\mathcal{F}(x)$. }
\label{F80}
\end{figure}

In Fig.~\ref{F80}, we compare the dependences $\Delta\rho_{xx}(B)$ measured on the structure 1 at $n=4.5\times 10^{11}$~cm$^{-2}$ [Fig.~\ref{F30}(a)]  and reduced to zero temperature with that calculated within the framework of the ``toy'' model with the use of the following parameters; $m=0.025\,m_0$ (the experimental value), $\gamma=1.5$~meV, $\alpha=5.7\times 10^{-6}$~meV$\cdot$cm, and $\mathcal{K} / \tau_{12}=1$ \footnote{The value of the parameter $\mathcal{K} / \tau_{12}=1$, at first glance, indicates that the model is inapplicable, because the mixing of states of the branches 1 and 2 is so great that separate oscillations from these branches should not be observed.  In fact, when $W_{ij}$ depends on $\Delta_{ij}$, the meaning  of the prefactor $\mathcal{K}$  introduced in Eq.~(\ref{eq100}) changes, it is not more equal to the average  relaxation time $(n_1\tau_1+n_2\tau_2)/(n_1+n_2)$.
}.  As seen this simple phenomenological model describes the experimental data quit  well. A similar agreement between the experimental and calculated dependences is observed for all the  structures under study.

At this point, it worth emphasizing that we do not believe that the values of the parameters $h$ and $b$ responsible in our ``toy'' model for antiresonance character of the $\Delta_{ij}$ dependence of $W_{ij}$  have a physical meaning. We just want to demonstrate that this dependence radically changes MISO. Obviously,  the microscopic model should be developed to analyze the data quantitatively.

\section{Conclusion}
\label{conlc}

The magneto-intersubband oscillations (MISO) have been experimentally studied in the  single quantum wells based on the gapless  semiconductor HgTe and on the conventional narrow-gap semiconductor In$_{0.75}$Ga$_{0.25}$As. The HgTe quantum well grown on substrates of different orientations with different quantum well widths corresponding both to  ``inverted'' and to ``normal'' band ordering are investigated. It has been experimentally established that  the electron energy spectrum in all these cases is split due to spin-orbit interaction, therewith the mutual positions of the MISO minima and antinodes of the SdH oscillations in the magnetic field are opposite to that observed in the double quantum wells or in the wide quantum wells with two subbands occupied. We assume  that the unusual mutual positions of the MISO extrema and the SdH oscillation antinodes originate from the dependence of the probability of transitions between the Landau levels of different branches on the difference in their energies. The ``toy'' model allowing us to take this effect into account is considerd. It has been  shown that even a slight decrease in the transition probability in the resonance of the Landau levels belonging to different split branches  leads to a significant change in the mutual position of the MISO and antinodes of the SdH oscillations and, at reasonable values of the parameters, gives a good agreement with the experimental dependences of the oscillations $\rho_{xx}(B)$. We have also  shown that even weak $\Delta_{ij}$ dependence of the transition probability changes  the magnetic field dependence of the MISO amplitude that should be taken into account under determining the quantum relaxation time from the magnetooscillation experiments.

\begin{acknowledgments}
We are grateful to A.~A. Bykov, I.~V. Gornyi, D.~G. Polyakov, O.~E. Raichev, M.~A. Zudov, and V.~Ya. Aleshkin for useful discussions. The work has been supported in part by the Russian Foundation for Basic
Research (Grant No. 18-02-00050), by  Act 211 Government of the Russian Federation (Agreement No.~02.A03.21.0006),  by  the Ministry of Science and Higher Education of the Russian Federation (Project No. FEUZ-2020-0054), and by the FASO of Russia (theme ``Electron'' No. 01201463326).
\end{acknowledgments}


\begin{thebibliography}{26}%
\makeatletter
\providecommand \@ifxundefined [1]{%
 \@ifx{#1\undefined}
}%
\providecommand \@ifnum [1]{%
 \ifnum #1\expandafter \@firstoftwo
 \else \expandafter \@secondoftwo
 \fi
}%
\providecommand \@ifx [1]{%
 \ifx #1\expandafter \@firstoftwo
 \else \expandafter \@secondoftwo
 \fi
}%
\providecommand \natexlab [1]{#1}%
\providecommand \enquote  [1]{``#1''}%
\providecommand \bibnamefont  [1]{#1}%
\providecommand \bibfnamefont [1]{#1}%
\providecommand \citenamefont [1]{#1}%
\providecommand \href@noop [0]{\@secondoftwo}%
\providecommand \href [0]{\begingroup \@sanitize@url \@href}%
\providecommand \@href[1]{\@@startlink{#1}\@@href}%
\providecommand \@@href[1]{\endgroup#1\@@endlink}%
\providecommand \@sanitize@url [0]{\catcode `\\12\catcode `\$12\catcode
  `\&12\catcode `\#12\catcode `\^12\catcode `\_12\catcode `\%12\relax}%
\providecommand \@@startlink[1]{}%
\providecommand \@@endlink[0]{}%
\providecommand \url  [0]{\begingroup\@sanitize@url \@url }%
\providecommand \@url [1]{\endgroup\@href {#1}{\urlprefix }}%
\providecommand \urlprefix  [0]{URL }%
\providecommand \Eprint [0]{\href }%
\providecommand \doibase [0]{http://dx.doi.org/}%
\providecommand \selectlanguage [0]{\@gobble}%
\providecommand \bibinfo  [0]{\@secondoftwo}%
\providecommand \bibfield  [0]{\@secondoftwo}%
\providecommand \translation [1]{[#1]}%
\providecommand \BibitemOpen [0]{}%
\providecommand \bibitemStop [0]{}%
\providecommand \bibitemNoStop [0]{.\EOS\space}%
\providecommand \EOS [0]{\spacefactor3000\relax}%
\providecommand \BibitemShut  [1]{\csname bibitem#1\endcsname}%
\let\auto@bib@innerbib\@empty
\bibitem [{\citenamefont {Lifshits}\ and\ \citenamefont
  {Kosevich}(1955)}]{LifKos55}%
  \BibitemOpen
  \bibfield  {author} {\bibinfo {author} {\bibfnamefont {I.~M.}\ \bibnamefont
  {Lifshits}}\ and\ \bibinfo {author} {\bibfnamefont {A.~M.}\ \bibnamefont
  {Kosevich}},\ }\href@noop {} {\bibfield  {journal} {\bibinfo  {journal} {Zh.
  Eksp. Teor. Fiz.}\ }\textbf {\bibinfo {volume} {29}},\ \bibinfo {pages} {730}
  (\bibinfo {year} {1955})},\ \translation{Sov. Phys. JETP \textbf{2}, 636
  (1956)}\BibitemShut {NoStop}%
\bibitem [{\citenamefont {Coleridge}\ \emph {et~al.}(1989)\citenamefont
  {Coleridge}, \citenamefont {Stoner},\ and\ \citenamefont
  {Fletcher}}]{Coleridge89}%
  \BibitemOpen
  \bibfield  {author} {\bibinfo {author} {\bibfnamefont {P.~T.}\ \bibnamefont
  {Coleridge}}, \bibinfo {author} {\bibfnamefont {R.}~\bibnamefont {Stoner}}, \
  and\ \bibinfo {author} {\bibfnamefont {R.}~\bibnamefont {Fletcher}},\ }\href
  {\doibase 10.1103/PhysRevB.39.1120} {\bibfield  {journal} {\bibinfo
  {journal} {Phys. Rev. B}\ }\textbf {\bibinfo {volume} {39}},\ \bibinfo
  {pages} {1120} (\bibinfo {year} {1989})}\BibitemShut {NoStop}%
\bibitem [{\citenamefont {Dmitriev}\ \emph {et~al.}(2012)\citenamefont
  {Dmitriev}, \citenamefont {Mirlin}, \citenamefont {Polyakov},\ and\
  \citenamefont {Zudov}}]{Dmitriev12}%
  \BibitemOpen
  \bibfield  {author} {\bibinfo {author} {\bibfnamefont {I.~A.}\ \bibnamefont
  {Dmitriev}}, \bibinfo {author} {\bibfnamefont {A.~D.}\ \bibnamefont
  {Mirlin}}, \bibinfo {author} {\bibfnamefont {D.~G.}\ \bibnamefont
  {Polyakov}}, \ and\ \bibinfo {author} {\bibfnamefont {M.~A.}\ \bibnamefont
  {Zudov}},\ }\href {\doibase 10.1103/RevModPhys.84.1709} {\bibfield  {journal}
  {\bibinfo  {journal} {Rev. Mod. Phys.}\ }\textbf {\bibinfo {volume} {84}},\
  \bibinfo {pages} {1709} (\bibinfo {year} {2012})}\BibitemShut {NoStop}%
\bibitem [{\citenamefont {Polyanovsky}(1988)}]{Polyanovs88}%
  \BibitemOpen
  \bibfield  {author} {\bibinfo {author} {\bibfnamefont {V.}~\bibnamefont
  {Polyanovsky}},\ }\href@noop {} {\bibfield  {journal} {\bibinfo  {journal}
  {Fiz. Tekh. Poluprovodn.}\ }\textbf {\bibinfo {volume} {22}},\ \bibinfo
  {pages} {2230} (\bibinfo {year} {1988})},\ \translation{Sov. Phys. Semicond.
  \textbf{22}, 1408 (1988)}\BibitemShut {NoStop}%
\bibitem [{\citenamefont {Leadley}\ \emph {et~al.}(1992)\citenamefont
  {Leadley}, \citenamefont {Fletcher}, \citenamefont {Nicholas}, \citenamefont
  {Tao}, \citenamefont {Foxon},\ and\ \citenamefont {Harris}}]{Leadly92}%
  \BibitemOpen
  \bibfield  {author} {\bibinfo {author} {\bibfnamefont {D.~R.}\ \bibnamefont
  {Leadley}}, \bibinfo {author} {\bibfnamefont {R.}~\bibnamefont {Fletcher}},
  \bibinfo {author} {\bibfnamefont {R.~J.}\ \bibnamefont {Nicholas}}, \bibinfo
  {author} {\bibfnamefont {F.}~\bibnamefont {Tao}}, \bibinfo {author}
  {\bibfnamefont {C.~T.}\ \bibnamefont {Foxon}}, \ and\ \bibinfo {author}
  {\bibfnamefont {J.~J.}\ \bibnamefont {Harris}},\ }\href {\doibase
  10.1103/PhysRevB.46.12439} {\bibfield  {journal} {\bibinfo  {journal} {Phys.
  Rev. B}\ }\textbf {\bibinfo {volume} {46}},\ \bibinfo {pages} {12439}
  (\bibinfo {year} {1992})}\BibitemShut {NoStop}%
\bibitem [{\citenamefont {Mamani}\ \emph {et~al.}(2008)\citenamefont {Mamani},
  \citenamefont {Gusev}, \citenamefont {Lamas}, \citenamefont {Bakarov},\ and\
  \citenamefont {Raichev}}]{Mamani08}%
  \BibitemOpen
  \bibfield  {author} {\bibinfo {author} {\bibfnamefont {N.~C.}\ \bibnamefont
  {Mamani}}, \bibinfo {author} {\bibfnamefont {G.~M.}\ \bibnamefont {Gusev}},
  \bibinfo {author} {\bibfnamefont {T.~E.}\ \bibnamefont {Lamas}}, \bibinfo
  {author} {\bibfnamefont {A.~K.}\ \bibnamefont {Bakarov}}, \ and\ \bibinfo
  {author} {\bibfnamefont {O.~E.}\ \bibnamefont {Raichev}},\ }\href {\doibase
  10.1103/PhysRevB.77.205327} {\bibfield  {journal} {\bibinfo  {journal} {Phys.
  Rev. B}\ }\textbf {\bibinfo {volume} {77}},\ \bibinfo {pages} {205327}
  (\bibinfo {year} {2008})}\BibitemShut {NoStop}%
\bibitem [{\citenamefont {Averkiev}\ \emph {et~al.}(2001)\citenamefont
  {Averkiev}, \citenamefont {Golub}, \citenamefont {Tarasenko},\ and\
  \citenamefont {Willander}}]{Averkiev01-1}%
  \BibitemOpen
  \bibfield  {author} {\bibinfo {author} {\bibfnamefont {N.~S.}\ \bibnamefont
  {Averkiev}}, \bibinfo {author} {\bibfnamefont {L.~E.}\ \bibnamefont {Golub}},
  \bibinfo {author} {\bibfnamefont {S.~A.}\ \bibnamefont {Tarasenko}}, \ and\
  \bibinfo {author} {\bibfnamefont {M.}~\bibnamefont {Willander}},\ }\href
  {\doibase 10.1088/0953-8984/13/11/309} {\bibfield  {journal} {\bibinfo
  {journal} {Journal of Physics: Condensed Matter}\ }\textbf {\bibinfo {volume}
  {13}},\ \bibinfo {pages} {2517} (\bibinfo {year} {2001})}\BibitemShut
  {NoStop}%
\bibitem [{\citenamefont {Raikh}\ and\ \citenamefont
  {Shahbazyan}(1994)}]{Raikh94}%
  \BibitemOpen
  \bibfield  {author} {\bibinfo {author} {\bibfnamefont {M.~E.}\ \bibnamefont
  {Raikh}}\ and\ \bibinfo {author} {\bibfnamefont {T.~V.}\ \bibnamefont
  {Shahbazyan}},\ }\href {\doibase 10.1103/PhysRevB.49.5531} {\bibfield
  {journal} {\bibinfo  {journal} {Phys. Rev. B}\ }\textbf {\bibinfo {volume}
  {49}},\ \bibinfo {pages} {5531} (\bibinfo {year} {1994})}\BibitemShut
  {NoStop}%
\bibitem [{\citenamefont {Raichev}(2010)}]{Raichev10}%
  \BibitemOpen
  \bibfield  {author} {\bibinfo {author} {\bibfnamefont {O.~E.}\ \bibnamefont
  {Raichev}},\ }\href {\doibase 10.1103/PhysRevB.81.195301} {\bibfield
  {journal} {\bibinfo  {journal} {Phys. Rev. B}\ }\textbf {\bibinfo {volume}
  {81}},\ \bibinfo {pages} {195301} (\bibinfo {year} {2010})}\BibitemShut
  {NoStop}%
\bibitem [{\citenamefont {Sander}\ \emph {et~al.}(1998)\citenamefont {Sander},
  \citenamefont {Holmes}, \citenamefont {Harris}, \citenamefont {Maude},\ and\
  \citenamefont {Portal}}]{Sander98}%
  \BibitemOpen
  \bibfield  {author} {\bibinfo {author} {\bibfnamefont {T.~H.}\ \bibnamefont
  {Sander}}, \bibinfo {author} {\bibfnamefont {S.~N.}\ \bibnamefont {Holmes}},
  \bibinfo {author} {\bibfnamefont {J.~J.}\ \bibnamefont {Harris}}, \bibinfo
  {author} {\bibfnamefont {D.~K.}\ \bibnamefont {Maude}}, \ and\ \bibinfo
  {author} {\bibfnamefont {J.~C.}\ \bibnamefont {Portal}},\ }\href {\doibase
  10.1103/PhysRevB.58.13856} {\bibfield  {journal} {\bibinfo  {journal} {Phys.
  Rev. B}\ }\textbf {\bibinfo {volume} {58}},\ \bibinfo {pages} {13856}
  (\bibinfo {year} {1998})}\BibitemShut {NoStop}%
\bibitem [{\citenamefont {Mamani}\ \emph
  {et~al.}(2009{\natexlab{a}})\citenamefont {Mamani}, \citenamefont {Gusev},
  \citenamefont {Raichev}, \citenamefont {Lamas},\ and\ \citenamefont
  {Bakarov}}]{Mamani09}%
  \BibitemOpen
  \bibfield  {author} {\bibinfo {author} {\bibfnamefont {N.~C.}\ \bibnamefont
  {Mamani}}, \bibinfo {author} {\bibfnamefont {G.~M.}\ \bibnamefont {Gusev}},
  \bibinfo {author} {\bibfnamefont {O.~E.}\ \bibnamefont {Raichev}}, \bibinfo
  {author} {\bibfnamefont {T.~E.}\ \bibnamefont {Lamas}}, \ and\ \bibinfo
  {author} {\bibfnamefont {A.~K.}\ \bibnamefont {Bakarov}},\ }\href {\doibase
  10.1103/PhysRevB.80.075308} {\bibfield  {journal} {\bibinfo  {journal} {Phys.
  Rev. B}\ }\textbf {\bibinfo {volume} {80}},\ \bibinfo {pages} {075308}
  (\bibinfo {year} {2009}{\natexlab{a}})}\BibitemShut {NoStop}%
\bibitem [{\citenamefont {Mamani}\ \emph
  {et~al.}(2009{\natexlab{b}})\citenamefont {Mamani}, \citenamefont {Gusev},
  \citenamefont {da~Silva}, \citenamefont {Raichev}, \citenamefont {Quivy},\
  and\ \citenamefont {Bakarov}}]{Mamani09-1}%
  \BibitemOpen
  \bibfield  {author} {\bibinfo {author} {\bibfnamefont {N.~C.}\ \bibnamefont
  {Mamani}}, \bibinfo {author} {\bibfnamefont {G.~M.}\ \bibnamefont {Gusev}},
  \bibinfo {author} {\bibfnamefont {E.~C.~F.}\ \bibnamefont {da~Silva}},
  \bibinfo {author} {\bibfnamefont {O.~E.}\ \bibnamefont {Raichev}}, \bibinfo
  {author} {\bibfnamefont {A.~A.}\ \bibnamefont {Quivy}}, \ and\ \bibinfo
  {author} {\bibfnamefont {A.~K.}\ \bibnamefont {Bakarov}},\ }\href {\doibase
  10.1103/PhysRevB.80.085304} {\bibfield  {journal} {\bibinfo  {journal} {Phys.
  Rev. B}\ }\textbf {\bibinfo {volume} {80}},\ \bibinfo {pages} {085304}
  (\bibinfo {year} {2009}{\natexlab{b}})}\BibitemShut {NoStop}%
\bibitem [{\citenamefont {Wiedmann}\ \emph {et~al.}(2010)\citenamefont
  {Wiedmann}, \citenamefont {Gusev}, \citenamefont {Raichev}, \citenamefont
  {Bakarov},\ and\ \citenamefont {Portal}}]{Wiedmann10}%
  \BibitemOpen
  \bibfield  {author} {\bibinfo {author} {\bibfnamefont {S.}~\bibnamefont
  {Wiedmann}}, \bibinfo {author} {\bibfnamefont {G.~M.}\ \bibnamefont {Gusev}},
  \bibinfo {author} {\bibfnamefont {O.~E.}\ \bibnamefont {Raichev}}, \bibinfo
  {author} {\bibfnamefont {A.~K.}\ \bibnamefont {Bakarov}}, \ and\ \bibinfo
  {author} {\bibfnamefont {J.~C.}\ \bibnamefont {Portal}},\ }\href {\doibase
  10.1103/PhysRevB.82.165333} {\bibfield  {journal} {\bibinfo  {journal} {Phys.
  Rev. B}\ }\textbf {\bibinfo {volume} {82}},\ \bibinfo {pages} {165333}
  (\bibinfo {year} {2010})}\BibitemShut {NoStop}%
\bibitem [{\citenamefont {Coleridge}(1990)}]{Col90}%
  \BibitemOpen
  \bibfield  {author} {\bibinfo {author} {\bibfnamefont {P.~T.}\ \bibnamefont
  {Coleridge}},\ }\href {http://stacks.iop.org/0268-1242/5/i=9/a=006}
  {\bibfield  {journal} {\bibinfo  {journal} {Semicond. Sci. Technol.}\
  }\textbf {\bibinfo {volume} {5}},\ \bibinfo {pages} {961} (\bibinfo {year}
  {1990})}\BibitemShut {NoStop}%
\bibitem [{\citenamefont {Bykov}\ \emph {et~al.}(2019)\citenamefont {Bykov},
  \citenamefont {Strygin}, \citenamefont {Goran}, \citenamefont {Marchishin},
  \citenamefont {Nomokonov}, \citenamefont {Bakarov}, \citenamefont {Abedi},\
  and\ \citenamefont {Vitkalov}}]{Bykov19}%
  \BibitemOpen
  \bibfield  {author} {\bibinfo {author} {\bibfnamefont {A.~A.}\ \bibnamefont
  {Bykov}}, \bibinfo {author} {\bibfnamefont {I.~S.}\ \bibnamefont {Strygin}},
  \bibinfo {author} {\bibfnamefont {A.~V.}\ \bibnamefont {Goran}}, \bibinfo
  {author} {\bibfnamefont {I.~V.}\ \bibnamefont {Marchishin}}, \bibinfo
  {author} {\bibfnamefont {D.~V.}\ \bibnamefont {Nomokonov}}, \bibinfo {author}
  {\bibfnamefont {A.~K.}\ \bibnamefont {Bakarov}}, \bibinfo {author}
  {\bibfnamefont {S.}~\bibnamefont {Abedi}}, \ and\ \bibinfo {author}
  {\bibfnamefont {S.~A.}\ \bibnamefont {Vitkalov}},\ }\href@noop {} {\bibfield
  {journal} {\bibinfo  {journal} {Pis'ma Zh. Eksp. Teor. Fiz.}\ }\textbf
  {\bibinfo {volume} {109}},\ \bibinfo {pages} {401} (\bibinfo {year}
  {2019})},\ \translation{JETP Letters \textbf{109}, 400 (2019)}\BibitemShut
  {NoStop}%
\bibitem [{\citenamefont {Minkov}\ \emph {et~al.}(2019)\citenamefont {Minkov},
  \citenamefont {Rut}, \citenamefont {Sherstobitov}, \citenamefont
  {Dvoretski},\ and\ \citenamefont {Mikhailov}}]{Minkov19}%
  \BibitemOpen
  \bibfield  {author} {\bibinfo {author} {\bibfnamefont {G.}~\bibnamefont
  {Minkov}}, \bibinfo {author} {\bibfnamefont {O.}~\bibnamefont {Rut}},
  \bibinfo {author} {\bibfnamefont {A.}~\bibnamefont {Sherstobitov}}, \bibinfo
  {author} {\bibfnamefont {S.}~\bibnamefont {Dvoretski}}, \ and\ \bibinfo
  {author} {\bibfnamefont {N.}~\bibnamefont {Mikhailov}},\ }\href@noop {}
  {\bibfield  {journal} {\bibinfo  {journal} {Pis'ma Zh. Eksp. Teor. Fiz.}\
  }\textbf {\bibinfo {volume} {110}},\ \bibinfo {pages} {274} (\bibinfo {year}
  {2019})},\ \translation{JETP Letters \textbf{110}, 301 (2019)}\BibitemShut
  {NoStop}%
\bibitem [{\citenamefont {Gerchikov}\ and\ \citenamefont
  {Subashiev}(1990)}]{Gerchikov90}%
  \BibitemOpen
  \bibfield  {author} {\bibinfo {author} {\bibfnamefont {L.~G.}\ \bibnamefont
  {Gerchikov}}\ and\ \bibinfo {author} {\bibfnamefont {A.}~\bibnamefont
  {Subashiev}},\ }\href@noop {} {\bibfield  {journal} {\bibinfo  {journal}
  {Phys. Stat. Sol. (b)}\ }\textbf {\bibinfo {volume} {160}},\ \bibinfo {pages}
  {443} (\bibinfo {year} {1990})}\BibitemShut {NoStop}%
\bibitem [{\citenamefont {Zhang}\ \emph {et~al.}(2001)\citenamefont {Zhang},
  \citenamefont {Pfeuffer-Jeschke}, \citenamefont {Ortner}, \citenamefont
  {Hock}, \citenamefont {Buhmann}, \citenamefont {Becker},\ and\ \citenamefont
  {Landwehr}}]{Zhang01}%
  \BibitemOpen
  \bibfield  {author} {\bibinfo {author} {\bibfnamefont {X.~C.}\ \bibnamefont
  {Zhang}}, \bibinfo {author} {\bibfnamefont {A.}~\bibnamefont
  {Pfeuffer-Jeschke}}, \bibinfo {author} {\bibfnamefont {K.}~\bibnamefont
  {Ortner}}, \bibinfo {author} {\bibfnamefont {V.}~\bibnamefont {Hock}},
  \bibinfo {author} {\bibfnamefont {H.}~\bibnamefont {Buhmann}}, \bibinfo
  {author} {\bibfnamefont {C.~R.}\ \bibnamefont {Becker}}, \ and\ \bibinfo
  {author} {\bibfnamefont {G.}~\bibnamefont {Landwehr}},\ }\href {\doibase
  10.1103/PhysRevB.63.245305} {\bibfield  {journal} {\bibinfo  {journal} {Phys.
  Rev. B}\ }\textbf {\bibinfo {volume} {63}},\ \bibinfo {pages} {245305}
  (\bibinfo {year} {2001})}\BibitemShut {NoStop}%
\bibitem [{\citenamefont {Novik}\ \emph {et~al.}(2005)\citenamefont {Novik},
  \citenamefont {Pfeuffer-Jeschke}, \citenamefont {Jungwirth}, \citenamefont
  {Latussek}, \citenamefont {Becker}, \citenamefont {Landwehr}, \citenamefont
  {Buhmann},\ and\ \citenamefont {Molenkamp}}]{Novik05}%
  \BibitemOpen
  \bibfield  {author} {\bibinfo {author} {\bibfnamefont {E.~G.}\ \bibnamefont
  {Novik}}, \bibinfo {author} {\bibfnamefont {A.}~\bibnamefont
  {Pfeuffer-Jeschke}}, \bibinfo {author} {\bibfnamefont {T.}~\bibnamefont
  {Jungwirth}}, \bibinfo {author} {\bibfnamefont {V.}~\bibnamefont {Latussek}},
  \bibinfo {author} {\bibfnamefont {C.~R.}\ \bibnamefont {Becker}}, \bibinfo
  {author} {\bibfnamefont {G.}~\bibnamefont {Landwehr}}, \bibinfo {author}
  {\bibfnamefont {H.}~\bibnamefont {Buhmann}}, \ and\ \bibinfo {author}
  {\bibfnamefont {L.~W.}\ \bibnamefont {Molenkamp}},\ }\href {\doibase
  10.1103/PhysRevB.72.035321} {\bibfield  {journal} {\bibinfo  {journal} {Phys.
  Rev. B}\ }\textbf {\bibinfo {volume} {72}},\ \bibinfo {pages} {035321}
  (\bibinfo {year} {2005})}\BibitemShut {NoStop}%
\bibitem [{\citenamefont {Bernevig}\ \emph {et~al.}(2006)\citenamefont
  {Bernevig}, \citenamefont {Hughes},\ and\ \citenamefont
  {Zhang}}]{Bernevig06}%
  \BibitemOpen
  \bibfield  {author} {\bibinfo {author} {\bibfnamefont {B.~A.}\ \bibnamefont
  {Bernevig}}, \bibinfo {author} {\bibfnamefont {T.~L.}\ \bibnamefont
  {Hughes}}, \ and\ \bibinfo {author} {\bibfnamefont {S.-C.}\ \bibnamefont
  {Zhang}},\ }\href {\doibase 10.1126/science.1133734} {\bibfield  {journal}
  {\bibinfo  {journal} {Science}\ }\textbf {\bibinfo {volume} {314}},\ \bibinfo
  {pages} {1757} (\bibinfo {year} {2006})}\BibitemShut {NoStop}%
\bibitem [{\citenamefont {Zholudev}(2013)}]{ZholudevPhD}%
  \BibitemOpen
  \bibfield  {author} {\bibinfo {author} {\bibfnamefont {M.}~\bibnamefont
  {Zholudev}},\ }\href@noop {} {Ph.D. thesis},\ \bibinfo  {school} {University
  Montpellier 2, France} (\bibinfo {year} {2013})\BibitemShut {NoStop}%
\bibitem [{Note1()}]{Note1}%
  \BibitemOpen
  \bibinfo {note} {It should be clarified here. The terms \protect \textit
  {inverted} and \protect \textit {normal} with respect to the spectrum of
  quantum wells of gapless semiconductors differ from the similar terms
  \protect \textit {inverted} and \protect \textit {normal order} of the bands
  of bulk semiconductors. In the bulk semiconductors, the term \protect \textit
  {normal} means that the upper in energy is the doubly degenerate $\Gamma _6$
  band (conduction band), while the valence band is formed by the fourfold
  degenerate $\Gamma _8$ states. The term \protect \textit {inverted} means
  that the $\Gamma _8$ band is located higher in energy than the $\Gamma _6$
  band and one of its branches forms the conduction band. Therefore, regarding
  the spectrum of quantum wells, we use the quotation marks: ``normal'' and
  ``inverted''.}\BibitemShut {Stop}%
\bibitem [{\citenamefont {Langenbuch}\ \emph {et~al.}(2004)\citenamefont
  {Langenbuch}, \citenamefont {Suhrke},\ and\ \citenamefont
  {R\"ossler}}]{Langenbuch04}%
  \BibitemOpen
  \bibfield  {author} {\bibinfo {author} {\bibfnamefont {M.}~\bibnamefont
  {Langenbuch}}, \bibinfo {author} {\bibfnamefont {M.}~\bibnamefont {Suhrke}},
  \ and\ \bibinfo {author} {\bibfnamefont {U.}~\bibnamefont {R\"ossler}},\
  }\href {\doibase 10.1103/PhysRevB.69.125303} {\bibfield  {journal} {\bibinfo
  {journal} {Phys. Rev. B}\ }\textbf {\bibinfo {volume} {69}},\ \bibinfo
  {pages} {125303} (\bibinfo {year} {2004})}\BibitemShut {NoStop}%
\bibitem [{\citenamefont {Novokshonov}(2013)}]{Novoksh}%
  \BibitemOpen
  \bibfield  {author} {\bibinfo {author} {\bibfnamefont {S.~G.}\ \bibnamefont
  {Novokshonov}},\ }\href {\doibase 10.1063/1.4803177} {\bibfield  {journal}
  {\bibinfo  {journal} {Low Temperature Physics}\ }\textbf {\bibinfo {volume}
  {39}},\ \bibinfo {pages} {378} (\bibinfo {year} {2013})}\BibitemShut
  {NoStop}%
\bibitem [{\citenamefont {Mikhailov}\ \emph {et~al.}(2006)\citenamefont
  {Mikhailov}, \citenamefont {Smirnov}, \citenamefont {Dvoretsky},
  \citenamefont {Sidorov}, \citenamefont {Shvets}, \citenamefont {Spesivtsev},\
  and\ \citenamefont {Rykhlitski}}]{Mikhailov06}%
  \BibitemOpen
  \bibfield  {author} {\bibinfo {author} {\bibfnamefont {N.~N.}\ \bibnamefont
  {Mikhailov}}, \bibinfo {author} {\bibfnamefont {R.~N.}\ \bibnamefont
  {Smirnov}}, \bibinfo {author} {\bibfnamefont {S.~A.}\ \bibnamefont
  {Dvoretsky}}, \bibinfo {author} {\bibfnamefont {Y.~G.}\ \bibnamefont
  {Sidorov}}, \bibinfo {author} {\bibfnamefont {V.~A.}\ \bibnamefont {Shvets}},
  \bibinfo {author} {\bibfnamefont {E.~V.}\ \bibnamefont {Spesivtsev}}, \ and\
  \bibinfo {author} {\bibfnamefont {S.~V.}\ \bibnamefont {Rykhlitski}},\ }\href
  {\doibase 10.1504/IJNT.2006.008725} {\bibfield  {journal} {\bibinfo
  {journal} {Int. J. Nanotechnology}\ }\textbf {\bibinfo {volume} {3}},\
  \bibinfo {pages} {120 } (\bibinfo {year} {2006})}\BibitemShut {NoStop}%
\bibitem [{Note2()}]{Note2}%
  \BibitemOpen
  \bibinfo {note} {The value of the parameter $\protect \mathcal {K} / \tau
  _{12}=1$, at first glance, indicates that the model is inapplicable, because
  the mixing of states of the branches 1 and 2 is so great that separate
  oscillations from these branches should not be observed. In fact, when
  $W_{ij}$ depends on $\Delta _{ij}$, the meaning of the prefactor $\protect
  \mathcal {K}$ introduced in Eq.~(\ref {eq100}) changes, it is not more equal
  to the average relaxation time $(n_1\tau _1+n_2\tau
  _2)/(n_1+n_2)$.}\BibitemShut {Stop}%
\end{thebibliography}
%

\end{document}